\documentclass[10pt,a4paper]{article}

\usepackage[]{tgpagella}
\usepackage{amssymb,amsmath}
\usepackage{ifxetex,ifluatex}
\usepackage{fixltx2e} 
\ifnum 0\ifxetex 1\fi\ifluatex 1\fi=0 
  \usepackage[T1]{fontenc}
  \usepackage[utf8]{inputenc}
\else 
  \ifxetex
    \usepackage{mathspec}
  \else
    \usepackage{fontspec}
  \fi
  \defaultfontfeatures{Ligatures=TeX,Scale=MatchLowercase}
\fi
\IfFileExists{upquote.sty}{\usepackage{upquote}}{}
\IfFileExists{microtype.sty}{%
\usepackage{microtype}
\UseMicrotypeSet[protrusion]{basicmath} 
}{}
\usepackage[margin=2.25cm]{geometry}
\usepackage{hyperref}
\hypersetup{unicode=true,
            pdftitle={A numerically stable algorithm for integrating Bayesian models using Markov melding},
            pdfauthor={Andrew A. Manderson and Robert J. B. Goudie},
            pdfborder={0 0 0},
            breaklinks=true}
\usepackage{natbib}
\usepackage{graphicx,grffile}
\makeatletter
\def\maxwidth{\ifdim\Gin@nat@width>\linewidth\linewidth\else\Gin@nat@width\fi}
\def\maxheight{\ifdim\Gin@nat@height>\textheight\textheight\else\Gin@nat@height\fi}
\makeatother
\setkeys{Gin}{width=\maxwidth,height=\maxheight,keepaspectratio}
\IfFileExists{parskip.sty}{%
\usepackage{parskip}
}{
\setlength{\parindent}{0pt}
\setlength{\parskip}{6pt plus 2pt minus 1pt}
}
\ifx\paragraph\undefined\else
\let\oldparagraph\paragraph
\renewcommand{\paragraph}[1]{\oldparagraph{#1}\mbox{}}
\fi
\ifx\subparagraph\undefined\else
\let\oldsubparagraph\subparagraph
\renewcommand{\subparagraph}[1]{\oldsubparagraph{#1}\mbox{}}
\fi

\let\rmarkdownfootnote\footnote%
\def\footnote{\protect\rmarkdownfootnote}

\usepackage{titling}

  \title{A numerically stable algorithm for integrating Bayesian models using
Markov melding}
    \pretitle{\vspace{\droptitle}\centering\huge}
  \posttitle{\par}
    \author{Andrew A. Manderson\footnote{MRC Biostatistics Unit, University of
  Cambridge, Cambridge, United Kingdom and The Alan Turing Institute,
  British Library, London, United Kingdom. \emph{email:}
  \href{mailto:andrew.manderson@mrc-bsu.cam.ac.uk}{\nolinkurl{andrew.manderson@mrc-bsu.cam.ac.uk}}} \\ and \\Robert J. B. Goudie\footnote{MRC Biostatistics Unit, University of
  Cambridge, Cambridge, United Kingdom}}
    \preauthor{\centering\large\emph}
  \postauthor{\par}
      \predate{\centering\large\emph}
  \postdate{\par}
    \date{23 September, 2021}

\usepackage{amsfonts}
\usepackage{mathtools}

\usepackage{xurl}
\usepackage{tikz}
\usetikzlibrary{positioning, shapes, intersections, through, backgrounds, fit, decorations.pathmorphing}

\usepackage{longtable}
\usepackage{booktabs}
\usepackage{caption}
\usepackage{accents}

\usepackage{microtype}

\newlength{\dhatheight}
\newcommand{\doublehat}[1]{%
    \settoheight{\dhatheight}{\ensuremath{\widehat{#1}}}%
    \addtolength{\dhatheight}{-0.35ex}%
    \widehat{\vphantom{\rule{1pt}{\dhatheight}}%
    \smash{\widehat{#1}}}}

\newcommand{\pd}{\text{p}}
\newcommand{\q}{\text{q}}
\newcommand{\w}{\text{w}}
\newcommand{\pdr}{\text{r}}
\newcommand{\pdrh}{\widehat{\text{r}}}

\newcommand{\ppoolphi}{\pd_{\text{pool}}(\phi)}
\newcommand{\s}{\text{s}}
\newcommand{\tofs}{\text{t}}
\newcommand{\m}{\text{m}}

\newcommand{\ddest}{\text{s}}

\newcommand{\Nx}{N}

\newcommand{\Nw}{W}
\newcommand{\Nm}{M}

\newcommand{\phinu}{\phi_{\text{nu}}}
\newcommand{\phide}{\phi_{\text{de}}}

\newcommand{\wfindex}{w}
\newcommand{\sampleindex}{n}
\newcommand{\modelindex}{m}

\begin{document}
\maketitle

\begin{abstract}
\noindent
  When statistical analyses consider multiple data sources, Markov melding provides a method for combining the source-specific Bayesian models.
  Markov melding joins together submodels that have a common quantity.
  One challenge is that the prior for this quantity can be implicit, and its prior density must be estimated.
  We show that error in this density estimate makes the two-stage Markov chain Monte Carlo sampler employed by Markov melding unstable and unreliable.
  We propose a robust two-stage algorithm that estimates the required prior marginal self-density ratios using weighted samples, dramatically improving accuracy in the tails of the distribution.
  The stabilised version of the algorithm is pragmatic and provides reliable inference.
  We demonstrate our approach using an evidence synthesis for inferring HIV prevalence, and an evidence synthesis of A/H1N1 influenza.
  \\ \\
  \textit{keywords:} biased sampling; data integration; evidence synthesis; kernel density estimation; multi-source inference; self-density ratio; weighted sampling.
\end{abstract}

\section{Introduction}

Many modern applied statistical analyses consider several data sources,
which differ in size and complexity. The wide variety of problems and
information sources has produced numerous methods for multi-source
inference \citep{lanckriet:etal:04, yates:etal:17, besbeas:morgan:19},
as well as general methodologies including evidence synthesis methods
\citep{sutton:abrams:01, ades_multiparameter_2006, spiegelhalter_bayesian_2004}, and the data fusion model
\citep{kedem:oliveira:sverchkov:17}. These methods require an
appropriate joint model for all data, which can be challenging
to specify.

An alternative approach is to model smaller, simpler aspects of the
data, such that designing these \textit{submodels} is easier, then
combine the submodels. The premise is that the combination of many
smaller submodels will serve as a good approximation to a larger joint
model, which may be methodologically or computationally infeasible.
\textit{Markov melding} \citep{goudie:etal:18} is a methodology for
coherently combining these submodels. Specifically, Markov melding joins
together submodels that share a common quantity \(\phi\) into a single
joint model. Consider \(\Nm\) submodels indexed by
\(\modelindex = 1, \ldots, \Nm\) that share \(\phi\), have submodel
specific parameters \(\psi_{\modelindex}\) and submodel specific data
\(Y_{\modelindex}\), denoting the \(m\textsuperscript{th}\) submodel
\(\pd_{\modelindex}(\phi, \psi_{\modelindex}, Y_{\modelindex})\). Markov
melding forms a single joint \textit{melded model}
\(\pd_{\text{meld}}(\phi, \psi_{1}, \ldots, \psi_{\Nm}, Y_{1}, \ldots, Y_{\Nm})\),
which enables information to flow through \(\phi\) from one model to
another. The melded model posterior thus incorporates uncertainty from all
sources of data.

Multi-stage sampling methods are useful, pragmatic tools for estimating complex
joint models -- such as $\pd_{\text{meld}}$ -- in a computationally feasible
manner, and have been applied in settings including statistical genetics and
phylogeny \citep{tom:etal:10}, meta-analysis
\citep{lunn:etal:13,blomstedt:etal:19}, spatial statistics
\citep{hooten:johnson:brost:19}, and joint models in survival analysis
\citep{mauff:etal:20}. Whilst it is preferable to sample the joint posterior
directly, this is often infeasible due to the complexity of the model, the
size of the data, the limitations of probabilistic programming languages such
as \texttt{JAGS} and \texttt{Stan}, or the complications of re-expressing
complicated submodels in a common programming language
\citep{johnson:brost:hooten:20}. Improving the stability of multi-stage
estimation techniques is thus of interest to applied statisticians.

Evidence synthesis models consider multiple sources of data (evidence), including randomised controlled trials or observational studies, to understand complex phenomena.
Each source of data has an associated submodel and set of parameters it informs; combining all the submodels requires assuming deterministic or probabilistic relationships between the submodel-specific parameters.
For example, \citet{de_angelis_estimation_2014} collected many surveys of partially overlapping subpopulations, albeit at different frequencies, and combined these in an evidence synthesis model to estimate human immunodeficiency virus (HIV) prevalence in the United Kingdom.
An introduction to evidence synthesis can be found in Chapter 8 of \citet{spiegelhalter_bayesian_2004}; other applications include estimating the prevalence of campylobacteriosis \citep{albert_bayesian_2011} and influenza \citep{presanis:etal:14}.

We can form evidence synthesis models by applying Markov melding to the various sources of data and their submodels.
However, the common quantity \(\phi\) may be a complex, non-invertible function of the parameters in one of the submodels.
This is a challenge for Markov melding, as the method requires the prior marginal density of \(\phi\) under
each submodel \(\pd_{\modelindex}(\phi)\), which may not be analytically
tractable. Instead, prior samples of \(\phi\) are drawn, and the prior marginal
density is estimated using a kernel density estimate (KDE)
\(\widehat{\pd}_{\modelindex}(\phi)\) \citep{wand:jones:95}. However, the use
of a KDE in lieu of the analytic density function has poor implications for the
numerical stability of the Markov Chain Monte Carlo (MCMC) method used to
estimate the melded posterior, even in our low dimensional examples.
Specifically, we illustrate that the multi-stage MCMC sampler of
\citet{goudie:etal:18} is sensitive to error in
\(\widehat{\pd}_{\modelindex}(\phi)\), particularly in low probability regions.

To address this sensitivity, we first note that Markov melding strictly
only requires an estimate of the \textit{self-density ratio}
\citep{hiraoka:hamada:hori:14},
\(\pdr(\phinu, \phide) = \pd_{\modelindex}(\phinu) \mathop{/} \pd_{\modelindex}(\phide)\),
as we will show in Section~\ref{sec:markov-melding}. In
Section~\ref{sec:self-density-ratio-estimation} we
develop methodology that reduces the error in the self-density ratio
estimate \(\pdrh(\phinu, \phide)\) by using weighted-sample KDEs
\citep{vardi:85, jones:91}, which are more accurate in low probability
regions. Multiple weighted-sample estimates of \(\pdrh(\phinu, \phide)\)
are combined via a weighted average to further improve performance. We
call this methodology
\textit{weighted-sample self-density ratio estimation} (WSRE), and
demonstrate the effectiveness of our methodology in two examples. The
first is a toy example from \citet{ades:cliffe:02}.
We show that output from the multi-stage
estimation process that uses WSRE is closer to reference samples
than the \textit{naive} approach, which uses a single KDE for
\(\widehat{\pd}_{\modelindex}(\phi)\). The second example is an involved
evidence synthesis, previously considered in \citet{goudie:etal:18}.
Here we show that the multi-stage estimation process that employs WSRE
produces plausible samples, whilst the naive approach produces
nonsensical results.
In these examples \(\phi\) is a 1 or 2 dimensional quantity.
We discuss the applicability of our method for higher dimensional \(\phi\) in Section~\ref{sec:discussion}.

\section{Markov Melding}
\label{sec:markov-melding}

The Markov melding framework is able to join together any number of
submodels which share a common component \(\phi\). As the examples in this
paper only consider two submodels, we limit our exposition to the
\(\Nm = 2\) model case; for the more general case see
\citet{goudie:etal:18}. Markov melding constructs a joint model using
the conditional distributions for submodel-specific parameters $\psi_{\modelindex}$ and data $Y_{\modelindex}$, denoted $\pd_{\modelindex}(\psi_{\modelindex}, Y_{\modelindex} \mid \phi)$.
These conditional distributions are then combined with a global prior for
\(\phi\) called the \textit{pooled prior} \(\ppoolphi\), which we
discuss in Section~\ref{sec:forming-the-pooled-prior}.
Mathematically, assuming that the supports of the relevant conditional,
joint, and marginal distributions containing \(\phi\) are appropriate,
we define the \textit{melded joint distribution} as
\begin{align}
  \pd_{\text{meld}}(\phi, \psi_{1}, \psi_{2}, Y_{1}, Y_{2})
  &=
    \ppoolphi
    \pd_{1}(\psi_{1}, Y_{1} \mid  \phi)
    \pd_{2}(\psi_{2}, Y_{2} \mid  \phi)
    \label{eqn:melded-joint-1} \\
  &=
    \ppoolphi
    \frac{\pd_{1}(\phi, \psi_{1}, Y_{1})}{\pd_{1}(\phi)}
    \frac{\pd_{2} (\phi, \psi_{2}, Y_{2})}{\pd_{2}(\phi)}.
    \label{eqn:melded-joint-2}
\end{align}
The submodel-specific conditional densities
\(\pd_{\modelindex}(\psi_{\modelindex}, Y_{\modelindex}) \mid \phi)\) may
be analytically intractable.
Hence, it is easier to work with the submodel-joint densities
$\pd_{\modelindex}(\phi, \psi_{\modelindex}, Y_{\modelindex})$ and their prior
marginal distributions $\pd_{\modelindex}(\phi)$ as specified in
Equation~\eqref{eqn:melded-joint-2}, because the former can be factorised into
the data generating process specified during submodel construction.

\subsection{Forming the pooled prior}
\label{sec:forming-the-pooled-prior}

The pooled prior should represent previous knowledge of \(\phi\) in the absence of other information.
A general approach to constructing \(\ppoolphi\) is to consider a
weighted combination of the prior marginal distributions
\(\pd_{\modelindex}(\phi)\), with submodel weights
\(\lambda_{\modelindex}\).
Selection of the pooling method and specific values of the weights is a topic
covered in much detail elsewhere \citep{clemen:winkler:99, ohagan:etal:06}; a
full summary of this field is beyond the scope of this article.
For the examples considered in this paper we
form \(\ppoolphi\) via logarithmic pooling:
\(\ppoolphi \propto \pd_{1}(\phi)^{\lambda_{1}}\pd_{2}(\phi)^{\lambda_{2}}\),
with \(\lambda_{1} = \lambda_{2} = \frac{1}{2}\). Logarithmic pooling
also allows us to use the methodology we develop in
Section~\ref{sec:self-density-ratio-estimation} in the
pooled prior.

\subsection{Two-stage Markov chain Monte Carlo sampler}
\label{sec:multi-stage-markov-chain-monte-carlo-sampler}

Directly estimating the melded model's posterior distribution
\begin{align}
  \pd_{\text{meld}} (\phi, \psi_{1}, \psi_{2} \mid Y_{1}, Y_{2})
    \, & \propto \,
    \ppoolphi
    \frac{\pd_{1}(\phi, \psi_{1}, Y_{1})}{\pd_{1}(\phi)}
    \frac{\pd_{2} (\phi, \psi_{2}, Y_{2})}{\pd_{2}(\phi)}.
    \label{eqn:melded-joint-two}
\end{align}
necessitates simultaneously evaluating both submodels.
This can be impractical if the submodels are implemented in different
probabilistic programming languages or have bespoke implementations.
We use the two-stage Markov chain Monte Carlo
(MCMC) sampler of \cite{goudie:etal:18} to sample from
\(\pd_{\text{meld}}(\phi, \psi_{1}, \psi_{2} \mid Y_{1}, Y_{2})\) without the need to
evaluate Equation~\eqref{eqn:melded-joint-two} all at once.
This involves a two-stage MCMC procedure, first sampling from a partial product of the terms in
Equation~\eqref{eqn:melded-joint-two}, then using these
samples as a proposal distribution in the second stage. The result is a
convenient cancellation of the common terms in the stage two acceptance
probability, whilst still ensuring that the final samples come from the
melded posterior distribution of Equation~\eqref{eqn:melded-joint-two}.

In stage one of the sampler we may, for example, opt to target the first
submodel \(\pd_{1}\), but with an (improper) flat prior for \(\phi\)
\begin{equation*}
  \pd_{\text{meld}, 1} (\phi, \psi_{1} \mid Y_{1}) \propto
  \frac {
    \pd_{1}(\phi, \psi_{1}, Y_{1})
  } {
    \pd_{1}(\phi)
  },
  \label{eqn:stage-one-target}
\end{equation*} so we construct a
standard Markov chain in which a proposed move from
\((\phi, \psi_{1}) \rightarrow (\phi^{*}, \psi_{1}^{*})\), with proposal
density \(\q(\phi^{*}, \psi_{1}^{*} \mid \phi, \psi_{1})\), is accepted
with probability
\begin{align}
  \alpha((\phi^{*}, \psi_{1}^{*}), (\phi, \psi_{1}))
  &= \frac {
    \pd_{1}(\phi^{*}, \psi_{1}^{*}, Y_{1}) 
    \pd_{1} (\phi)
    \q(\phi, \psi_{1} \mid \phi^{*}, \psi_{1}^{*})
  } {
    \pd_{1}(\phi, \psi_{1}, Y_{1}) 
    \pd_{1} (\phi^{*})
    \q(\phi^{*}, \psi_{1}^{*} \mid \phi, \psi_{1})
  }.
  \label{eqn:stage-one-acceptance}
\end{align} This
Markov chain asymptotically emits samples from \(\pd_{\text{meld}, 1}\).

In stage two we update \(\phi\) and \(\psi_{2}\) using
Metropolis-within-Gibbs updates, targeting the full melded posterior
distribution of Equation~\eqref{eqn:melded-joint-two}.
Updating \(\phi\) uses the stage one samples as a proposal distribution.
For a sample of size \(\Nx\) from \(\pd_{\text{meld}, 1}\) denoted
\(\{\phi_{\sampleindex}^{(\text{meld}, 1)}\}_{\sampleindex = 1}^{\Nx}\) we sample an
index \(\sampleindex^{*}\) uniformly at random between 1 and \(\Nx\),
and use the corresponding value as the proposal
\(\phi^{*} = \phi_{\sampleindex^{*}}^{(\text{meld}, 1)}\). This results in a stage two
acceptance probability for a move from \(\phi \rightarrow \phi^{*}\) of
\begin{align}
\begin{split}
  \alpha (\phi^{*}, \phi)
  &= \frac {
    \pd_{\text{pool}} (\phi^{*})
    \pd_{1}(\phi^{*}, \psi_{1}, Y_{1})
    \pd_{2}(\phi^{*}, \psi_{2}, Y_{2})
    \pd_{1}(\phi)
    \pd_{2}(\phi)
  } {
    \pd_{\text{pool}} (\phi)
    \pd_{1}(\phi, \psi_{1}, Y_{1})
    \pd_{2}(\phi, \psi_{2}, Y_{2})
    \pd_{1}(\phi^{*})
    \pd_{2}(\phi^{*})
  } \frac {
    \pd_{1}(\phi, \psi_{1}, Y_{1}) \pd_{1}(\phi^{*})
  } {
    \pd_{1}(\phi^{*}, \psi_{1}, Y_{1}) \pd_{1}(\phi)
  }
  \end{split}
  \nonumber \\ 
  &= 
  \frac {
    \pd_{\text{pool}} (\phi^{*})
    \pd_{2}(\phi^{*}, \psi_{2}, Y_{2})
    \pd_{2}(\phi)
  } {
    \pd_{\text{pool}} (\phi)
    \pd_{2}(\phi, \psi_{2}, Y_{2})
    \pd_{2}(\phi^{*})
  },
  \label{eqn:stage-two-acceptance-two}
\end{align} since all
stage one terms cancel, providing a form of ``modularisation'' in the
algorithm. The update for \(\psi_{2}\) has an acceptance probability for
a move from \(\psi_{2} \rightarrow \psi_{2}^{*}\), drawn from a proposal
distribution \(\q(\psi_{2}^{*} \mid \psi_{2})\), of
\begin{equation*}
  \alpha(\psi_{2}^{*}, \psi_{2}) = 
  \frac {
    \pd_{2} (\phi, \psi_{2}^{*}, Y_{2})
  } { 
    \pd_{2} (\phi, \psi_{2}, Y_{2})
  }
  \frac {
    \q (\psi_{2} \mid \psi_{2}^{*})
  } {
    \q (\psi_{2}^{*} \mid \psi_{2})
  },
  \label{eqn:stage-two-psi-acceptance}
\end{equation*} as
all terms that do not contain \(\psi_{2}\) cancel. Samples from the
melded posterior distribution for \(\psi_{1}\),
\(\pd_{\text{meld}}(\psi_{1} \mid Y_{1}, Y_{2})\), can be obtained by
storing the indices \(\sampleindex\) used to draw values of \(\phi\)
from \(\{\phi_{\sampleindex}^{(\text{meld}, 1)}\}_{\sampleindex = 1}^{\Nx}\) in stage two.
The stored indices are then used to resample the stage one samples $\{\psi_{1, n}^{(\text{meld}, 1)}\}_{n = 1}^{\Nx}$ yielding samples from \(\pd_{\text{meld}}(\psi_{1} \mid Y_{1}, Y_{2})\).

An interesting property of
Equations~\eqref{eqn:stage-one-acceptance}~and~\eqref{eqn:stage-two-acceptance-two}
is that our interaction with the unknown prior marginal distribution is
limited to the \textit{self-density ratio}
\(\pdr(\phi, \phi^{*}) = \pd_{\modelindex}(\phi) \mathop{/} \pd_{\modelindex}(\phi^{*})\).
In Section~\ref{sec:self-density-ratio-estimation} we
develop methodology that uses self-density ratios to improve the numerical stability of the acceptance probability calculations.

We do not have to target
\(\pd_{\text{meld}, 1} (\phi, \psi_{1} \mid Y_{1})\) with an improper prior in stage one; we
are free to choose any of the components of
Equation~\eqref{eqn:melded-joint-two}. The choice of
stage one components will affect MCMC mixing, yet is often constrained by the
practicalities of sampling the subposterior distributions. In the
example of Section~\ref{sec:h1n1-example-output} the common quantity $\phi$ is a non-invertible function of parameters in $\pd_{1}$, and it is
possible to sample from the subposterior
\(\pd_{1}(\phi, \psi_{1} \mid Y_{1})\) using \texttt{JAGS}. Hence, we
draw stage one samples from \(\pd_{1}(\phi, \psi_{1} \mid Y_{1})\), with
stage two, implemented partially in \texttt{Stan}, accounting for the
remaining terms: \(1 \mathop{/} \pd_{1}(\phi)\),
\(\pd_{2}(\phi, \psi_{2} \mid Y_{2}) \mathop{/} \pd_{2}(\phi)\), and
\(\ppoolphi\). This process highlights another interesting advantage of
Markov melding; we can use samples produced from one statistical
software package in combination with a model implemented in another,
mixing and matching as is most convenient.

\subsection{Naive prior marginal estimation}
\label{sec:interest-in-the-prior-marginal}

The expressions in
Equation~\eqref{eqn:stage-one-acceptance} and
Equation~\eqref{eqn:stage-two-acceptance-two} explicitly
include both models' prior marginal distributions
\(\pd_{\modelindex}(\phi)\) for \(\modelindex = 1, 2\), and implicitly
includes them in \(\ppoolphi\). In our examples we do not have analytic
expressions for these marginals. More generally, if $\phi$ is not a root node
in the directed acyclic graph representation of either submodel (see e.g.
$\pi_{12}$ in Figure~\ref{fig:model-dag}), or is the aggregate output of a
non-invertible deterministic link function, then the analytic form of
$\pd_{\modelindex}(\phi)$ will likely be intractable.

The approach proposed by \citet{goudie:etal:18}, which we call the
\textit{naive} approach, estimates the prior marginal distributions by
sampling
\(\pd_{\modelindex}(\phi, \psi_{\modelindex}, Y_{\modelindex})\) for
each model using simple Monte Carlo, as the samples of \(\phi\) will be
distributed according to the correct marginal, and employs a standard
KDE \(\widehat{\pd}_{\modelindex}(\phi)\) \citep{wand:jones:95}. The
two-stage sampler then targets the corresponding estimate of the
melded posterior
\begin{align}
  \hat{\pd}_{\text{meld}}(\phi, \psi_{1}, \psi_{2} \mid Y_{1}, Y_{2})
  \, & \propto \,
  \hat{\pd}_{\text{pool}}(\phi)
  \frac{\pd_{1}(\phi, \psi_{1}, Y_{1})}{\hat{\pd}_{1}(\phi)}
  \frac{\pd_{2} (\phi, \psi_{2}, Y_{2})}{\hat{\pd}_{2}(\phi)},
  \label{eqn:melded-joint-two-kde}
\end{align}
where $\hat{\pd}_{\text{pool}}(\phi)$ is the approximation to $\ppoolphi$ obtained by plugging in $\hat{\pd}_{\modelindex}(\phi)$ for $\modelindex = 1, 2$.

\subsection{Numerical issues in the naive approach}
\label{sec:numerical-issues-in-the-naive-approach}

Sampling the melded posterior using Equation~\eqref{eqn:melded-joint-two-kde}
can be numerically unstable.
Say we propose a move from \(\phi \rightarrow \phi^{*}\), where \(\phi^{*}\)
is particularly improbable under \(\pd_{\modelindex}\). The KDE
estimate at this value, \(\widehat{\pd}_{\modelindex}(\phi^{*})\), is
poor in terms of relative error
\begin{equation*}
  \left\|
    \frac {
      \widehat{\pd}_{\modelindex} (\phi^{*}) - \pd_{\modelindex} (\phi^{*}) 
    } {
      \pd_{\modelindex} (\phi^{*})
    }
  \right\|_{1},
  \label{eqn:absolute-relative-error}
\end{equation*}
particularly in the tails of the distribution
\citep{koekemoer:swanepoel:08}. In our experience, the KDE is typically
an underestimate in the tails, which can lead to an explosion in the
self-density ratio estimate \(\hat{\pdr}(\phi, \phi^{*}) = \hat{\pd}_{\modelindex}(\phi) \mathop{/} \hat{\pd}_{\modelindex}(\phi^{*})\).
Hence, improbable values for \(\phi^{*}\)
are accepted far too often. Once at this improbable value,
i.e.~when \(\phi\) is improbable under
\(\pd_{\modelindex}(\phi)\), the error in the KDE then leads to a
dramatically reduced value for the acceptance probability. This results
in Markov chains that get stuck at improbable values. For example, see
the top left panel of
Figure~\ref{fig:stage_two_phi_traces}.

In which stage this instability arises depends on which prior marginal densities are intractable, and how the terms in Equation \eqref{eqn:melded-joint-two} are apportioned across the stages.
In the example of Section~\ref{sec:hiv-example-output}, $\pd_{1}(\phi)$ is unknown and is part of both stage one (in Equation \eqref{eqn:stage-one-acceptance}) and stage two (via $\pd_{\text{pool}}(\phi)$ in Equation \eqref{eqn:stage-two-acceptance-two}).
Thus both stages are numerically brittle.
Our second example, contained in Section~\ref{sec:h1n1-example-output}, represents a more typical scenario, where the first submodel posterior is used as the proposal for the melded posterior.
In this case, all unknown prior marginal terms are factorised into the stage two target, and the instability is confined to the second stage.

\section{Self-density ratio estimation}
\label{sec:self-density-ratio-estimation}

As described in Section~\ref{sec:numerical-issues-in-the-naive-approach}, the self-density ratios associated with both $\pd_{1}(\phi)$ and $\pd_{2}(\phi)$ may be required by the two-stage MCMC algorithm.
To simplify notation, we consider in this section a generic joint density $\pd(\phi, \gamma)$ that we can evaluate pointwise, but whose marginal $\pd(\phi) = \int \pd(\phi, \gamma)\text{d}\gamma$ we cannot obtain analytically.
Our interest is in the \textit{self-density ratio} evaluated at $\phinu$ and $\phide$ (the subscripts are abbreviations of numerator and denominator respectively) which we denote as
\begin{equation*}
  \pdr(\phinu, \phide) = \frac{
    \pd(\phinu)
  } {
    \pd(\phide)
  }
  = \frac {
    \int \pd(\phinu, \gamma) \text{d} \gamma
  } {
    \int \pd(\phide, \gamma) \text{d} \gamma
  }.
  \label{eqn:self-density-ratio-def}
\end{equation*}
In our examples we set $\phinu = \phi$ and $\phide = \phi^{*}$ for use in Equations \eqref{eqn:stage-one-acceptance} and \eqref{eqn:stage-two-acceptance-two}; and define $\gamma = (\psi_{\modelindex}, Y_{\modelindex})$ and $\pd = \pd_{\modelindex}$ where $\modelindex = 1$ or $2$ as appropriate (see Sections \ref{sec:hiv-example-output} and \ref{sec:h1n1-example-output} for details).

To avoid the numerical issues associated with the
naive approach, we need to improve the ratio estimate
\(\pdrh(\phinu, \phide)\) for improbable values of \(\phinu\) and
\(\phide\), e.g.~values more than two standard
deviations away from the mean. The fundamental flaw in the naive
approach in this context is that it minimises the absolute error in
the high density region (HDR) of \(\pd(\phi)\), i.e.~the
region \(R_{\varepsilon}(\pd(\phi)) = \{\phi : \pd(\phi) > \varepsilon\}\).
But this is not necessarily the sole region of interest, and we are
concerned with minimising the relative error. To address this we
reweight \(\pd(\phi)\) towards a particular region, and thus obtain a
more accurate estimate in that region. We then exploit the fact that we
only interact with the prior marginal distribution via its self-density
ratio to combine estimates from multiple reweighted distributions.

\subsection{Single weighting function}
\label{sec:single-weighting-function-case}

We can shift \(\pd(\phi)\) by multiplying the joint distribution
\(\pd(\phi, \gamma)\) by a known weighting function \(\w(\phi;\xi)\),
controlled by parameter \(\xi\), then account for this shift in our KDE.
This will improve the accuracy of the KDE in the region to which we
shift the marginal. We first generate \(\Nx\) samples denoted
\(\{(\phi_{\sampleindex}, \gamma_{\sampleindex})\}_{\sampleindex = 1}^{\Nx}\),
from a weighted version of the joint
distribution
\begin{equation}
  \{(\phi_{\sampleindex}, \gamma_{\sampleindex})\}_{\sampleindex = 1}^{\Nx} 
    \, \sim \,
    \frac{1}{Z_{1}}
    \pd(\phi, \gamma) \w(\phi; \xi),
  \label{eqn:weighted-joint-def}
\end{equation}
where $Z_{1} = \iint \pd(\phi, \gamma) \w(\phi; \xi) \text{d}\phi \text{d}\gamma$ is the normalising constant.
The samples
\(\{\phi_{\sampleindex}^{(\s)}\}_{\sampleindex = 1}^{\Nx}\), obtained by ignoring the samples of $\gamma_{\sampleindex}$, are distributed
according to a weighted version \(\s(\phi; \xi)\) of the marginal
distribution \(\pd(\phi)\)
\begin{equation*}
  \{\phi_{\sampleindex}^{(\s)}\}_{\sampleindex = 1}^{\Nx} \, \sim \,
    \frac{1}{Z_{2}}
    \pd(\phi)\w(\phi ; \xi) = \s(\phi; \xi),
\end{equation*}
where $Z_{2} = \int \pd(\phi) \w(\phi; \xi) \text{d}\phi$.
Typically \eqref{eqn:weighted-joint-def} cannot be sampled by simple Monte Carlo; instead we employ MCMC.

Using the samples \(\{\phi_{\sampleindex}^{(\s)}\}_{\sampleindex = 1}^{\Nx}\)
from \(\s(\phi; \xi)\) we compute a weighted kernel density estimate
\citep{jones:91}, with bandwidth \(h\), kernel \(\text{K}_{h}\), and
normalising constant \(Z_{3}\)
\begin{equation}
  \doublehat{\pd}(\phi) = 
    \frac{1} {Z_{3} \Nx h}
    \sum_{\sampleindex = 1}^{\Nx}
    (\w(\phi_{\sampleindex}; \xi))^{-1} \text{K}_{h}(\phi - \phi_{\sampleindex}^{(\s)}),
  \label{eqn:jones-estimator-def}
\end{equation} and form our
weighted-sample self-density ratio estimate
\begin{equation*}
  \doublehat{\pdr}(\phinu, \phide) = \frac{
    \doublehat{\pd}(\phinu)
  } {
    \doublehat{\pd}(\phide)
  }
  = \frac {
    \sum_{\sampleindex = 1}^{\Nx}
    (\w(\phi_{\sampleindex}; \xi))^{-1} \text{K}_{h}(\phinu - \phi_{\sampleindex}^{(\s)})
  } {
    \sum_{\sampleindex = 1}^{\Nx}
    (\w(\phi_{\sampleindex}; \xi))^{-1} \text{K}_{h}(\phide - \phi_{\sampleindex}^{(\s)})
  }
  .
  \label{eqn:ratio-estimate-def}
\end{equation*} The cancellation of
the normalisation constant \(Z_{3}\) is crucial, as accurately
estimating constants like \(Z_{3}\) is known to be challenging.

\subsection{Choice of weighting function}
\label{sec:restrictions-on-the-choice-of-weighting-function}

The choice of \(\w(\phi;\xi)\) affects both the validity and efficacy of our methodology.
The weighted marginal \(\s(\phi; \xi)\) must satisfy the requirements for a density for our method to be valid.
Hence, the specific form of \(\w(\phi;\xi)\) is subject to some restrictions.
Our first requirement is that \(\w(\phi;\xi) > 0\) for all \(\phi\) in the support of \(\pd(\phi, \gamma)\).
We also require that the weighted joint distribution, defined in \eqref{eqn:weighted-joint-def}, has finite integral, to ensure that it can be normalised to a probability distribution, and that the marginal \(\s(\phi; \xi)\) is positive over the support of interest, also with finite integral.

\subsection{Multiple weighting functions}
\label{sec:multiple-weighting-functions}

The methodology of
Section~\ref{sec:single-weighting-function-case}
produces a single estimate \(\doublehat{\pdr}(\phinu, \phide)\) using
\(\doublehat{\pd}(\phi)\) from
Equation~\eqref{eqn:jones-estimator-def}. It is accurate
for values in the HDR of \(\s(\phi; \xi)\),
i.e.~\(R_{\varepsilon}(\s(\phi; \xi))\), and we can control the
location of \(R_{\varepsilon}(\s(\phi; \xi))\) through \(\xi\). This is similar to
importance sampling, with \(\s(\phi; \xi)\) acting as the proposal density.
\citet{Nakayama:11} notes importance sampling can be used to improve the
mean square error (MSE) of a KDE in a specific local region, at the cost
of an increase in global MSE. To ameliorate the decrease in global
performance, we specify multiple regions in which we want accurate
estimates for \(\doublehat{\pd}(\phi)\), and then combine the
corresponding estimates of \(\doublehat{\pdr}(\phinu, \phide)\) to
provide a single estimate that is accurate across all regions.

We elect to use \(\Nw\) different weighting functions, indexed by
\(\wfindex = 1, \ldots, \Nw\), with function-specific parameters
\(\xi_{\wfindex}\) denoted
\(\w(\phi;\xi_{\wfindex})\). Samples are then drawn from each
of the \(\Nw\) weighted distributions
\(\s_{\wfindex}(\phi; \xi_{\wfindex}) \propto \pd(\phi) \w(\phi;\xi_{\wfindex})\).
Denote the samples from the \(\wfindex\textsuperscript{th}\) weighted
distribution by
\(\{\phi_{\sampleindex}^{(\s_{\wfindex})}\}_{\sampleindex = 1}^{\Nx}\). Each
set of samples produces a separate ratio estimate
\(\doublehat{\pdr}_{\wfindex}(\phinu, \phide)\) in the manner described
in Section~\ref{sec:single-weighting-function-case}.

Each individual $\doublehat{\pdr}_{\wfindex}$ is accurate (in terms of relative accuracy) only in the HDR of $\s_{\wfindex}(\phi; \xi_{\wfindex})$.
Thus, when combining multiple ratio estimates, simply taking the mean of all $\wfindex = 1, \ldots, \Nw$ estimates (for a specific value of $\phinu$ and $\phide$) would not make use of our knowledge about the region in which $\doublehat{\pdr}_{\wfindex}$ is accurate.
We therefore propose a weighted average of all the individual ratio estimates, where the weights approximately come from $\s_{\wfindex}(\phinu; \xi_{\wfindex}) \s_{\wfindex}(\phide; \xi_{\wfindex})$ -- this quantity is largest when $\doublehat{\pdr}_{\wfindex}(\phinu, \phide)$ is most accurate.
This ensures the more accurate terms are given more weight in our final estimate.
Specifically, we use $\{\phi_{\sampleindex}^{(\s_{\wfindex})}\}_{\sampleindex = 1}^{\Nx}$ to compute a standard KDE of $\s_{\wfindex}(\phi; \xi_{\wfindex})$
\begin{equation*}
  \hat{\ddest}_{\wfindex}(\phi; \xi_{\wfindex}) = \frac{1}{\Nx h}
  \sum_{\sampleindex = 1}^{\Nx} \text{K}_{h}(\phi - \phi_{\sampleindex}^{(\s_{\wfindex})}).
  \label{eqn:ddest-direct-def}
\end{equation*}
Finally, we form the weighted-sample self-density ratio estimate $\doublehat{\pdr}_{\text{WSRE}}(\phinu, \phide)$, which is a weighted mean of the individual ratio estimates
\begin{equation*}
  \doublehat{\pdr}_{\text{WSRE}}(\phinu, \phide) =
  \frac {1} {Z_{4}}
    \sum_{\wfindex = 1}^{\Nw} \hat{\ddest}_{\wfindex}(\phinu, \phide; \xi_{\wfindex}) \doublehat{\pdr}_{\wfindex}(\phinu, \phide),
\end{equation*}
where \(\hat{\ddest}_{\wfindex} (\phinu, \phide; \xi_{\wfindex}) = \hat{\ddest}_{\wfindex} (\phinu; \xi_{\wfindex})\,\hat{\ddest}_{\wfindex} (\phide; \xi_{\wfindex})\) and $Z_{4} = \sum_{\wfindex = 1}^{\Nw} \hat{\ddest}_{\wfindex}(\phinu, \phide; \xi_{\wfindex})$.

\subsection{Choosing values for $\xi_{\wfindex}$}
\label{sec:choice-of-xi}

Consider a $D$-dimensional $\phi = (\phi^{[1]}, \ldots, \phi^{[D]})$ where $\phi^{[d]}$ is the $d$\textsuperscript{th} component of $\phi$, for $d = 1, \ldots, D$.
Assume we have a compact region of interest for the $d$\textsuperscript{th} component denoted $A_{d} = [a_{d}, b_{d}] \subseteq \text{supp}(\phi^{[d]})$, such that the overall region of interest $A$ can be defined as the Cartesian product of component-wise regions of interest $A = \bigtimes_{d = 1}^{D} A_{d}$.
We are interested in accurately evaluating the self-density ratio for two points in this region.
We will obtain $W$ choices for $\xi_{\wfindex}$ by specifying $V$ weighting functions for each of the $D$ components, such that $W = V^{D}$.

Assume that the weighting function $\w(\phi; \xi)$ is composed of $D$ independent component weighting functions
\begin{equation*}
  \w(\phi; \xi) = \prod_{d = 1}^{D} \m(\phi^{[d]}; \xi^{[d]}),
\end{equation*}
where $\xi^{[d]}$ is the $d$\textsuperscript{th} component of $\xi$.
We can then define the marginal of the weighted target
\begin{equation*}
    \tofs(\phi^{[d]}; \xi^{[d]})
    = \int \s(\phi; \xi) \text{d}\phi^{[-d]},
\end{equation*}
where $\phi^{[-d]}$ represents the $D - 1$ components of $\phi$ that are not $\phi^{[d]}$.
For typical choices of $\xi$ and $\w(\phi; \xi)$, the corresponding HDR of $\tofs(\phi^{[d]}; \xi^{[d]})$ does not span the region of
interest.
That is, $\lvert R_{\varepsilon}(\tofs(\phi^{[d]}; \xi^{[d]})) \rvert \ll \lvert A_{d} \rvert$.

Our aim is to choose, for each of the $d$ components, values $v = 1, \ldots, V$ of $\xi^{[d]}$ denoted $\{\xi_{v, d}\}_{v = 1}^{V}$, yielding weighting functions $\m(\phi^{[d]}, \xi_{v, d})$ and corresponding $\tofs(\phi^{[d]}, \xi_{v, d})$, such that $\bigcup_{v = 1}^{V} R_{\varepsilon}(\tofs(\phi^{[d]}; \xi_{v, d})) \approx A_{d} = [a_{d}, b_{d}]$.
We employ the following heuristic argument, first choosing a ``minimum'' $\xi_{1, d}$ and a ``maximum'' $\xi_{V, d}$ such that
\begin{equation*}
  \xi_{1, d} \vcentcolon \, a_{d} \in R_{\varepsilon}(\tofs(\phi^{[d]}; \xi_{1, d})), \qquad
  \xi_{V, d} \vcentcolon \, b_{d} \in R_{\varepsilon}(\tofs(\phi^{[d]}; \xi_{V, d})).
\end{equation*}
In words, we choose a minimum value $\xi_{1, d}$ so that the corresponding HDR of the weighted target includes the lower limit of the region of interest.
An analogous argument is used to choose the maximum $\xi_{V, d}$.
We then interpolate $V - 2$ values between $\xi_{1, d}$ and $\xi_{V, d}$ ensuring that there is sufficient, but not complete, overlap between the corresponding HDRs.

Denote an element from the set of all $W$ possible values for the parameter of the weighting function with $\xi_{\wfindex} \in \bigtimes_{d = 1}^{D} \{\xi_{v, d}\}_{v = 1}^{V}$, noting that $\xi_{\wfindex}$ is a $D$-vector.

The practitioner typically has some knowledge of $p(\phi)$ and $A$ from prior predictive checks and previous attempts at running the two-stage sampler.
Thus only a small number of trial-and-error attempts should be needed to determine $\xi_{1, d}$ and $\xi_{V, d}$ for all dimensions.
These attempts are also used to check for overlap between the HDRs, and increase $V$ if the overlap is insufficient.
Section~\ref{sec:discussion} contains further discussion of this selection process and its relationship to umbrella sampling \citep{torrie_nonphysical_1977-1}

\subsection{Practicalities and software}
\label{software-implementation}

In our examples we use Gaussian density functions for \(\m(\phi^{[d]};\xi_{v, d})\),
\begin{equation*}
  \m(\phi^{[d]};\xi_{v, d}) =
    \frac{1} {\sqrt{2 \pi \sigma^{2}_{v, d}}}
    \text{exp} \left\{
      - \frac{1}{2 \sigma^{2}_{v, d}}
      (\phi - \mu_{v, d})^2
    \right\},
  \label{eqn:example-wf-intext}
\end{equation*}
with $\xi_{v, d} = (\mu_{v, d}, \sigma^{2}_{v, d})$, though we fix  $\sigma^{2}_{v, d} = \sigma^{2}_{d}$ for all $v$.
Our definition of sufficient overlap is that 0.95 empirical quantile of \(\tofs(\phi^{[d]}; \xi_{v, d})\) is equal or slightly greater than the 0.05 empirical quantile of \(\tofs(\phi^{[d]}; \xi_{v + 1, d})\), for $v = 1, \ldots, V - 1$.

Our implementation of our WSRE
methodology is available in an \texttt{R} \citep{rlang:19} package at
\url{https://github.com/hhau/wsre}. It is built on top of \texttt{Stan}
\citep{carpenter:etal:stan:17} and \texttt{Rcpp} \citep{rcpp:11}.
Package users supply a joint density \(\pd(\phi, \gamma)\) in the form
of a \texttt{Stan} model; choose the parameters \(\xi_{\wfindex}\) of
each of the \(\Nw\) weighting functions; and the number of samples
\(\Nx\) drawn from each \(\s_{\wfindex}(\phi; \xi_{\wfindex})\). The combined estimate
\(\doublehat{\pdr}_{\text{WSRE}}(\phinu, \phide)\) is returned. A vignette on
using \texttt{wsre} is included in the package, and documents the
specific form of \texttt{Stan} model required.

\section{An evidence synthesis for estimating the efficacy of HIV screening}
\label{sec:hiv-example-output}

To illustrate our approach we artificially split an existing joint model into two submodels, then compare the melded posterior estimates obtained by the two-stage algorithm using the naive and WSRE approaches.
Artificially splitting this joint model serves several purposes: it demonstrates that the numerical instability can occur in a simple, low dimensional setting; we can obtain a good parametric approximation to the prior marginal to use as a reference; and the simplicity of the model allows us to focus on our methodology, not the complexity of the model.

\subsection{Model}
\label{sec:hiv-model-def}

The model is a simple evidence synthesis model for inferring the efficacy of HIV screening in prenatal clinics \citep{ades:cliffe:02}, and has 8 \textit{basic parameters} $\rho_{1}, \rho_{2}, \ldots, \rho_{8}$, which are group membership probabilities for particular risk groups and subgroups thereof.
The first risk group partitions the prenatal clinic attendees into those born in sub-Saharan Africa (SSA), injecting drug users (IDU), and the remaining women.
These groups have corresponding membership probabilities $\rho_{1}, \rho_{2}$, and $1 - \rho_{1} - \rho_{2}$.
The groups are subdivided based on whether they are infected with HIV, with group specific HIV positivity $\rho_{3}, \rho_{4}$ and $\rho_{5}$ respectively; and if they had already been diagnosed before visiting the clinic, with pre-clinical diagnosis probabilities $\rho_{6}, \rho_{7}$ and $\rho_{8}$.
An additional probability is also included in the model, denoted $\rho_{9}$, which considers the prevalence of HIV serotype B.
This parameter enables the inclusion of study 12, which further informs the other basic parameters.
Table~\ref{tab:HIV-data} summarises the full joint model, including the $s = 1, \ldots, 12$ studies with observations $y_{s}$ and sample size $n_{s}$; the basic parameters $\rho_{1}, \ldots, \rho_{9}$; and the link functions that relate the study proportions $\pi_{1}, \ldots, \pi_{12}$ to the basic parameters.

\begin{table*}
\centering
\caption{\label{tab:HIV-data} HIV example: Study probabilities, their relationships to the basic parameters, and data used to inform the probabilities. For full details on the sources of the data see \citet{ades:cliffe:02}.}
\begin{tabular}{p{6cm}rrr}
\toprule
\em Parameter & \em Data & & \\ 
 & \em $y$ & \em $n$ & \em $y \mathop{/} n$ \\ 
\midrule
  $\pi_1 = \rho_{1}$ & 11,044 & 104,577 & 0.106 \\[8pt]
  $\pi_2 = \rho_{2}$ & 12 & 882 & 0.014 \\[8pt] 
  $\pi_3 = \rho_{3}$ & 252 & 15,428 & 0.016 \\[8pt] 
  $\pi_4 = \rho_{4}$ & 10 & 473 & 0.021 \\[8pt] 
  $\pi_5 = \frac{\rho_{4}\rho_{2} + \rho_{5}(1-\rho_{1} -\rho_{2})}{1-\rho_{1} }$ & 74 & 136,139 & 0.001 \\[8pt] 
  $\pi_6 = \rho_{3}\rho_{1}  + \rho_{4}\rho_{2} + \rho_{5}(1-\rho_{1} -\rho_{2})$ & 254 & 102,287 & 0.002 \\[8pt] 
  $\pi_7 = \frac{\rho_{6}\rho_{3}\rho_{1} }{\rho_{6}\rho_{3}\rho_{1}  + \rho_{7}\rho_{4}\rho_{2} + \rho_{8}\rho_{5}(1-\rho_{1} -\rho_{2})}$ & 43 & 60 & 0.717 \\[8pt] 
  $\pi_8 = \frac{\rho_{7}\rho_{4}\rho_{2}}{\rho_{7}\rho_{4}\rho_{2} + \rho_{8}\rho_{5}(1-\rho_{1} -\rho_{2})}$ & 4 & 17 & 0.235 \\[8pt] 
  $\pi_9 = \frac{\rho_{6}\rho_{3}\rho_{1} +\rho_{7}\rho_{4}\rho_{2}+\rho_{8}\rho_{5}(1-\rho_{1} -\rho_{2})}{\rho_{3}\rho_{1}  + \rho_{4}\rho_{2} + \rho_{5}(1-\rho_{1} -\rho_{2})}$ & 87 & 254 & 0.343 \\[8pt] 
  $\pi_{10} = \rho_{7}$ & 12 & 15 & 0.800 \\[8pt] 
  $\pi_{11} = \rho_{9}$ & 14 & 118 & 0.119 \\[8pt] 
  $\pi_{12} = \frac{\rho_{4}\rho_{2} + \rho_{9}\rho_{5}(1-\rho_{1} -\rho_{2})}{\rho_{4}\rho_{2} + \rho_{5}(1-\rho_{1} -\rho_{2})}$ & 5 & 31 & 0.161 \\ 
\bottomrule
\end{tabular}
\end{table*} 

We make one small modification to original model of \citet{ades:cliffe:02}, to better highlight the impact of WSRE on the melded posterior estimate.
The original model adopts a flat, Beta$(1, 1)$ prior for $\rho_{9}$.
This induces a prior on $\pi_{12}$ that is not flat, but not overly informative, making it difficult to demonstrate the issues caused by an inaccurate density estimate of the tail of the prior marginal distribution.
Instead, we adopt a Beta$(3, 1)$ prior for $\rho_{9}$.
This prior would have been reasonable for the time and place in which the original evidence synthesis was constructed, since the distribution of HIV serotypes differs considerably between North America and sub-Saharan Africa \citep{hemelaar:12}.

The code to reproduce this example is available at
\url{https://github.com/hhau/presanis-conflict-hiv-example}.

\subsection{Submodels formed by splitting}
\label{sec:model-particulars}

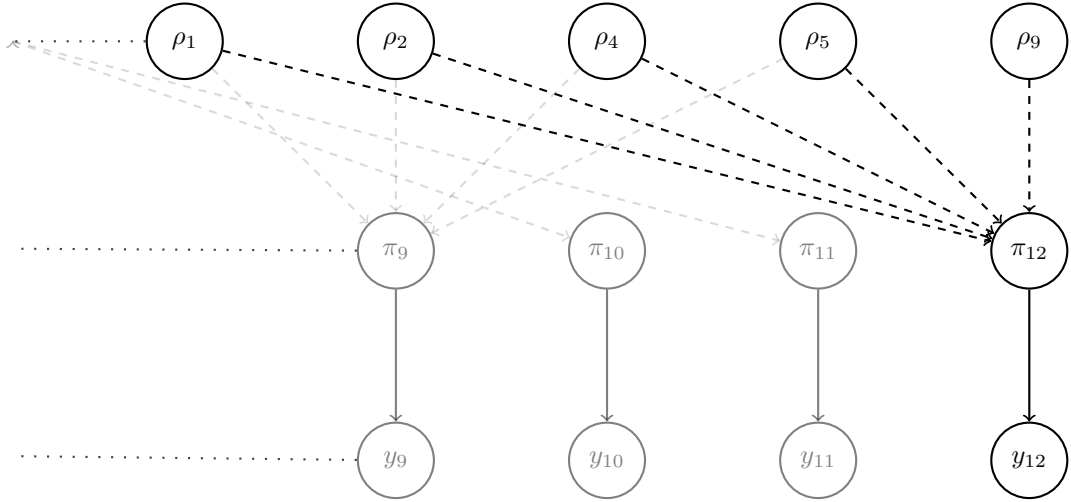
\begin{figure*}[!ht]
  \centering
  \begin{tikzpicture}[node distance = 1.75cm, thick, state/.style={circle, draw, minimum width = 1cm}]
    \node [state] (y12) {$y_{12}$};
    \node [state, opacity=0.5][left = of y12] (y11) {$y_{11}$};
    \node [state, opacity=0.5][left = of y11] (y10) {$y_{10}$};
    \node [state, opacity=0.5][left = of y10] (y9) {$y_{9}$};

    \node [state] [above = of y12] (p12) {$\pi_{12}$};
    \node [state, opacity=0.5] [left = of p12] (p11) {$\pi_{11}$};
    \node [state, opacity=0.5] [left = of p11] (p10) {$\pi_{10}$};
    \node [state, opacity=0.5] [left = of p10] (p9) {$\pi_{9}$};
    
    \node [state] [above = of p12] (w) {$\rho_{9}$};
    \node [state] [left = of w] (e) {$\rho_{5}$};
    \node [state] [left = of e] (d) {$\rho_{4}$};
    \node [state] [left = of d] (b) {$\rho_{2}$};
    \node [state] [left = of b] (a) {$\rho_{1}$};

    \coordinate [left = of a] (high);
    \coordinate [below =2.75cm of high] (mid);
    \coordinate [below =2.75cm of mid] (low);

    \draw [->] (p12) edge (y12);
    \draw [opacity=0.5] [->] (p11) edge (y11);
    \draw [opacity=0.5] [->] (p10) edge (y10);
    \draw [opacity=0.5][->] (p9) edge (y9);
    
    \draw [dashed] [->] (w) edge (p12);
    \draw [dashed] [->] (e) edge (p12);
    \draw [dashed] [->] (d) edge (p12);
    \draw [dashed] [->] (b) edge (p12);
    \draw [dashed] [->] (a) edge (p12);

    \draw [dashed, opacity=0.15] [->] (high) edge (p11);
    \draw [dashed, opacity=0.15] [->] (high) edge (p10);

    \draw [dashed, opacity=0.15] [->] (a) edge (p9);
    \draw [dashed, opacity=0.15] [->] (b) edge (p9);
    \draw [dashed, opacity=0.15] [->] (d) edge (p9);
    \draw [dashed, opacity=0.15] [->] (e) edge (p9);

    \draw [loosely dotted, opacity=0.75] (a) -- (high.center);
    \draw [loosely dotted, opacity=0.75] (p9) -- (mid.center);
    \draw [loosely dotted, opacity=0.75] (y9) -- (low.center);

  \end{tikzpicture}
  \caption{Partial directed acyclic graph (DAG) for the HIV model. The top row only depicts nodes that relate to $\pi_{12}$. Dashed lines indicate deterministic relationships between nodes, some of which are non-invertible. Solid lines indicate stochastic relationships.}
  \label{fig:model-dag}
\end{figure*}

In the full joint model study 12 informs the probability $\pi_{12}$, and provides indirect evidence for the basic parameters through the deterministic link function
\begin{equation*}
  \pi_{12} = \frac {
    \rho_{2}\rho_{4} + \rho_{9}\rho_{5}(1 - \rho_{1} - \rho_{2})
  } {
    \rho_{2}\rho_{4}  + \rho_{5}(1 - \rho_{1} - \rho_{2})
  }.
  \label{eqn:study-link-function}
\end{equation*}
Figure~\ref{fig:model-dag} is a DAG of the basic parameters in the full model that relate to $\pi_{12}$.
We consider splitting the model at the node corresponding to the expected proportion
$\pi_{12}$ in study 12, i.e.~we set the common quantity
$\phi = \pi_{12}$.

The first submodel ($\modelindex = 1$) considers data from studies 1
to 11 $Y_{1} = (y_{1}, \ldots, y_{11})$, corresponding study
proportions $(\pi_{1}, \ldots, \pi_{11})$, and all basic parameters
$\psi_{1} = (\rho_{1}, \ldots, \rho_{9})$. Note that the study
proportions are implicitly defined because they are deterministic
functions of the basic parameters. The joint distribution of this
submodel is
\begin{equation*}
  \pd_{1} (\rho_{1}, \ldots, \rho_{9}, y_{1}, \ldots, y_{12}) = 
    \pd(\rho_{1}) \ldots \pd(\rho_{9})
    \prod_{s = 1}^{11} \pd(y_{s} \mid \pi_{s}).
  \label{eqn:big-submodel}
\end{equation*}
The prior $\pd_{1}(\pi_{12})$ on the common quantity $\phi = \pi_{12}$
is implicitly defined, so its analytic form is unknown, hence it needs
to be estimated.

The second submodel ($\modelindex = 2$) pertains specifically to study
12, with data $Y_{2} = y_{12}$, the study 12 specific probability
$\phi = \pi_{12}$, and $\psi_{1} = \varnothing$. The joint distribution is
$\pd_{2}(\pi_{12}, y_{12}) = \pd_{2}(\pi_{12}) \pd(y_{12} \mid \pi_{12}).$
In more complex examples $\pd_{2}(\phi)$ may be implicitly defined,
and contribute substantially to the melded posterior. However, in this
simple example we are free to choose
$\pd_{2}(\pi_{12}) = \pd_{2}(\phi)$, and opt for a Beta$(1, 1)$
prior.

\subsection{Self-density ratio estimation}
\label{sec:example-details}

We now compute the self-density ratio estimate
\(\doublehat{\pdr}_{\text{WSRE}}(\phinu, \phide)\) of $\pd_{1}(\phinu) \mathop{/} \pd_{1}(\phide)$.
In the notation defined in Section~\ref{sec:choice-of-xi}, this example has $D = 1$, and we use $V = W = 7$ Gaussian weighting functions.
We fix the variance parameter of the weighting function \(\sigma_{\wfindex}^{2} = 0.08^2\) for all $\wfindex$, and use the heuristic described in Section~\ref{sec:choice-of-xi} to choose values for the mean parameter of the weighting function.
Specifically, we set the minimum to be $\xi_{1, 1} = \mu_{1} = 0.05$, with maximum $\xi_{7, 1} = \mu_{7} = 0.08$ and 5 additional, equally spaced values between these extrema.
We draw 3000 MCMC samples in total, apportioned equally across the $7$ weighting functions.
We thus draw 428 post warmup MCMC samples from each weighted target.

\begin{figure}
  \centering
  \includegraphics[width=\linewidth]{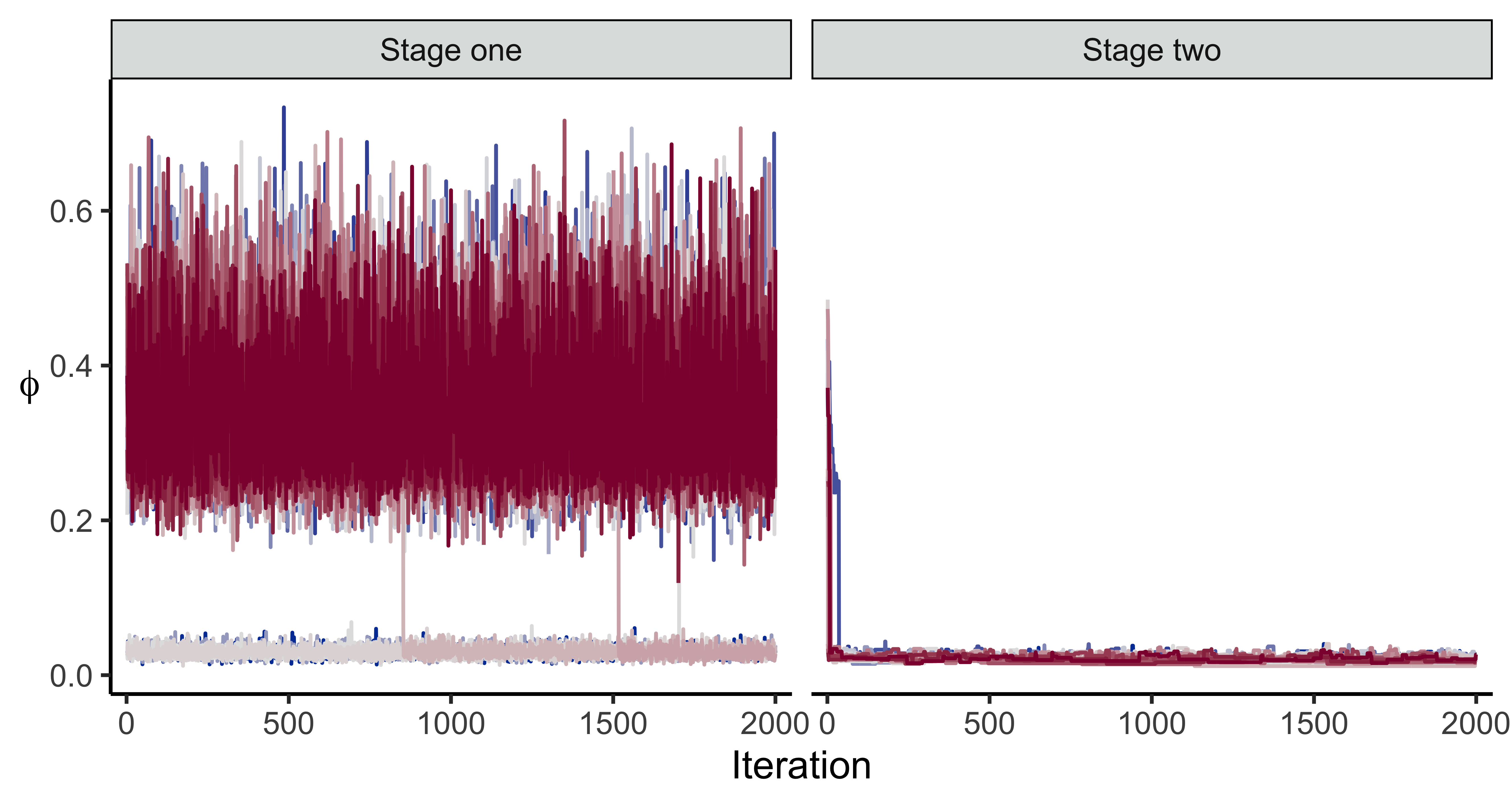}
  \caption{\textit{Left}: Stage one trace plot for $\phi$ using the naive method. At any moment in time chains can jump to the spurious mode, which is an artefact of $\hat{\pd}_{1}(\phi)$. \textit{right}: Corresponding stage two trace plot. The stage two target has the same numerical instability, and because the stage one samples are the proposal distribution, all chains encounter the instability.}
  \label{fig:stage_one_trace}
\end{figure}

\begin{figure}
  \centering
  \includegraphics[width=0.75\linewidth]{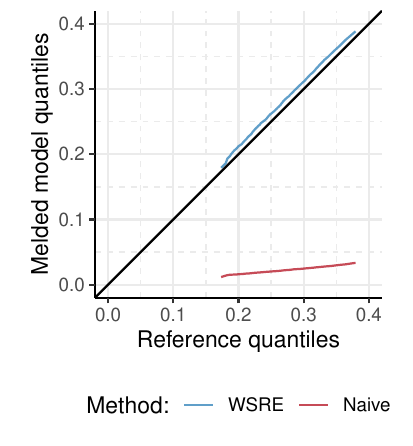}
  \caption{Quantile-quantile plot of the melded posterior quantiles using the WSRE approach (blue) and the naive approach (red). Both methods are comparable to the quantiles from the reference sample (x-axis).}
  \label{fig:qq_plot_phi}
\end{figure}

\subsection{Results}
\label{sec:hiv-example-results}

We compare the melded posterior obtained by the naive approach and using WSRE.
For a fair comparison, we estimate the prior marginal distribution of interest \(\widehat{\pd}_{1}(\phi)\) using 3000 Monte Carlo samples,
This set-up is slightly advantageous for the naive approach, which uses Monte Carlo samples, rather than the MCMC samples of the self-density ratio estimate; the naive approach makes use of a sample comprised of 3000 effective samples, whilst the self-density ratio estimate uses fewerthan 3000 effective samples.
A reference estimate of the melded posterior is
obtained using a parametric density estimate $\hat{\pd}_{\text{ref}, 1}(\phi)$
for the unknown prior marginal, based on $5 \times 10^{6}$ prior samples.
The reference sample also contains some error, as
$\hat{\pd}_{\text{ref}, 1}(\phi)$ is not perfect. However, in the absence
of an analytic form for $\pd_{1}(\phi)$ it serves as a very close approximation.
We estimate the melded posterior using the two-stage sampler of
Section~\ref{sec:multi-stage-markov-chain-monte-carlo-sampler},
targeting in stage one
\begin{equation}
  \pd_{\text{meld}, 1} (\phi, \psi_{1} \mid Y_{1}) \propto
  \frac {
      \pd_{1}(\phi, \psi_{1}, Y_{1})
  } {
      \pd_{1}(\phi)
  }.
  \label{eqn:stage_one_target}
\end{equation} and the full
melded posterior in stage two.
To demonstrate the numerical instability of interest, we run 24 chains that target $\pd_{\text{meld}, 1}$ in \eqref{eqn:stage_one_target} using the naive approach.

The left panel of Figure \ref{fig:stage_one_trace} displays the trace plot of the post-warmup samples.
Many chains have already converged to a spurious model around $\phi \approx 0.02$, and other chains jump to this mode after a variable number of additional iterations.
As discussed in Section \ref{sec:numerical-issues-in-the-naive-approach}, this mode is an artefact of the naive KDE employed for $\hat{\pd}_{1}(\phi)$, and is also visible in the corresponding stage two trace plot (right panel of Figure~\ref{fig:stage_one_trace}).
Because the stage one samples act as the proposal for stage two, all stage two chains quickly jump to the spurious mode.

The samples surrounding the spurious mode introduce substantial bias in estimate of the melded posterior under the naive approach.
This is visible in the quantile-quantile plot in Figure \ref{fig:qq_plot_phi}, where the naive approach produces an implausible estimate compared to the reference quantiles.
In contrast, the WSRE approach rectifies the numerical instability, and uses the two-stage sampler to produce a sensible estimate of the melded posterior.

\section{An evidence synthesis to estimate the severity of the H1N1 pandemic}
\label{sec:h1n1-example-output}

We now consider a more involved example, where the prior for the common
quantity does not have an analytical form under either submodel, and
the two priors contain a substantially different quantity of information.
\citet{presanis:etal:14} undertook a large evidence synthesis in order
to estimate the severity of the H1N1 pandemic amongst the population of
England. This model combines independent data on the number of suspected
influenza cases in hospital's intensive care unit (ICU) into a large
severity model. Here, we reanalyse the model introduced in
\citet{goudie:etal:18} that uses Markov melding to join the independent
ICU model (\(\modelindex = 1\)) with a simplified version of the larger,
complex severity model (\(\modelindex = 2\)). In this example the melded
model has no obvious implied joint model, so there are no simple ``gold
standard'' joint model estimates to use as a baseline reference.
However, we demonstrate that the naive approach is highly unstable,
whereas the WSRE approach produces stable results. The code to reproduce
all figures and outputs for this example is available at
\url{https://github.com/hhau/full-melding-example}.

\subsection{ICU submodel}
\label{sec:icu}

The data for the ICU submodel (\(\modelindex = 1\)) are aggregate weekly
counts of patients in the ICU of all the hospitals in England, for 78
days between December 2010 and February 2011. Observations were recorded
of the number of children \(a = 1\) and adults \(a = 2\) in the ICU on
days \(U = \{8, 15, \ldots, 78\}\), and we denote a specific weekly
observation as \(y_{a, t}\) for \(t \in U\).

To appropriately model the temporal nature of the weekly ICU data we use
a time inhomogeneous, thinned Poisson process with rate parameter
\(\lambda_{a, t}\) for \(t \in T\) where \(T = \{1, 2, \ldots, 78\}\).
This is the expected number of new ICU admissions; the corresponding age
group specific ICU exit rate is \(\mu_{a}\). There is also a discrepancy
between the observation times \(U\) and the daily support of our Poisson
process \(T\). We address this in the observation model
\begin{align}
  y_{a, t} &\sim \text{Pois}(\eta_{a, t}), \,\, t \in U, 
  \label{eqn:thinned-poisson-process-one} 
  \\
  \eta_{a, t} &= \sum_{u = 1}^{t} \lambda_{a, u} \exp\{-\mu_{a}(t - u)\}, \,\, t \in T,
  \label{eqn:thinned-poisson-process-two}
\end{align} through
different supports for \(t\) in
Equations~\eqref{eqn:thinned-poisson-process-one} and
\eqref{eqn:thinned-poisson-process-two}. An
identifiability assumption of \(\eta_{a, 1} = 0\) is required, which
enforces the reasonable assumption that no H1N1 influenza patients were
in the ICU at~time~\(t~=~0\).

Weekly virological positivity data are available at weeks
\(V = \{1, \ldots, 11\}\), and inform the proportion of influenza cases
which are attributable to the H1N1 virus \(\pi_{a, t}^{\text{pos}}\).
The virology data consists of the number of H1N1-positive swabs
\(z_{a, v}^{\text{pos}}\) and the total number of swabs
\(n_{a, z}^{\text{pos}}\) tested for influenza that week. This
proportion relates the counts to \(\pi_{a, t}^{\text{pos}}\) via a
truncated uniform prior~on~\(\pi_{a, t}^{\text{pos}}\),
\begin{align*}
  \pi_{a, t}^{\text{pos}} &\sim \text{Unif}(\omega_{a, v}, 1), \,\, t \in T
  \,\,
  \\ 
  z_{a, v}^{\text{pos}} &\sim \text{Bin}(n_{a, z}^{\text{pos}}, \omega_{a, v})
  \label{eqn:pipos-unif-model-two}, \,\, v \in V,\
\end{align*} with \(v = 1\) for
\(t = 1, 2, \ldots, 14\), and
\(v = \lfloor(t - 1) \mathop{/} 7 \rfloor\) for
\(t = 15, 16, \ldots, 78\) to align the temporal indices. The positivity
proportion \(\pi_{a, t}^{\text{pos}}\) is combined with
\(\lambda_{a, t}\) to compute the lower bound on the total number of
H1N1 cases
\(\phi_{a} = \sum\limits_{t \in T} \pi_{a, t}^{\text{pos}} \lambda_{a,t}\)
where \(\phi = (\phi_{1}, \phi_2)\) is the quantity common to both
submodels. This summation is a non-invertible function, which
necessitates either considering this model in stage one of our
two-stage sampler, or appropriately augmenting the definition of
\(\phi_{a}\) such that it is invertible. We elect to consider this
submodel in stage one, and further discuss model ordering in
Section~\ref{sec:discussion}.

Lastly, we specify priors for the remaining parameters. A lognormal
random-walk is used for the expected number of new admissions
\begin{align*}
  \log(\lambda_{a, 1}) \sim \text{Unif}(0, 250), \,&\, 
  \log(\lambda_{a, t}) \sim \text{N}(\log(\lambda_{a, t - 1}), \nu_{a}^{-2}), \\  
  \nu_{a} & \sim \text{Unif}(0.1, 2.7),
\end{align*}
 for
\(t = 2, 3, \ldots, 78\) and \(a = {1, 2}\). Age group specific exit
rates have informative priors \citep{presanis:etal:14}
\begin{equation*}
  \begin{gathered}
  \mu_{1} = \exp\{-\alpha\}, \,\,
  \mu_{2} = \exp\{-(\alpha + \beta)\}, \\  
  \alpha \sim \text{N}(2.7058, 0.0788^2), \,\, 
  \beta \sim \text{N}(-0.4969, 0.2048^2),
  \end{gathered}
  \label{eqn:prior-set-two}
\end{equation*} and the lower bound
on the positivity proportion has a flat prior,
\(\omega_{a, v} \sim \text{Unif}(0, 1)\), for \(v \in V\).

\subsection{Severity submodel}
\label{sec:severity}

A simplified version of the large severity model (\(\modelindex = 2\)) of
\citet{presanis:etal:14} is considered here, in
which parts of the severity model are collapsed into informative priors.
The cumulative number of ICU admissions \(\phi_{a}\) is assumed to be an
underestimate of the true number of ICU admissions due to H1N1,
\(\chi_{a}\). This motivates\\
\begin{align*}
  \phi_{a} \sim \text{Bin}(\chi_{a}, \pi^{\text{det}}), &\,\,
  \pi^{\text{det}} \sim \text{Beta}(6, 4),  \\
  \chi_{1} \sim \text{LN}(4.93, 0.17^2), &\,\,
  \chi_{2} \sim \text{LN}(7.71, 0.23^2),
  \label{eqn:sev-submodel}
\end{align*} where
\(\pi^{\text{det}}\) is the age group constant probability of detection,
and the priors on \(\chi_{a}\) appropriately summarise the remainder of
the large severity model.

\subsection{Prior distributions, stage one target}
\label{sec:prior-stage-one-comparison}

\begin{figure}
  \centering
  \includegraphics[width=\linewidth]{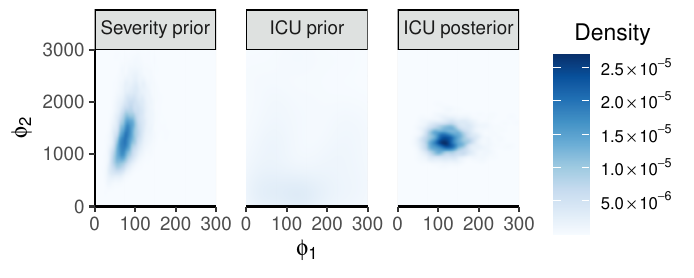}
  \caption{Heatmap of the severity submodel prior $\pd_{2}(\phi)$, ICU submodel prior $\pd_{1}(\phi)$, and the stage one (ICU submodel) posterior $\pd_{1}(\phi \mid Y_{1})$.}
  \label{fig:prior_comparison}
\end{figure}

Figure~\ref{fig:prior_comparison} displays
\(\pd_{\modelindex}(\phi)\) for both submodels, as well as the
subposterior for the ICU (\(\modelindex = 1\)) submodel
\(\pd_{1}(\phi \mid Y_{1})\). The melded posterior will be largely
influenced by the product of \(\pd_{1}(\phi \mid Y_{1})\) and
\(\pd_{2}(\phi)\), since \(\pd_{1}(\phi)\) is effectively uniform (see
the centre panel of Figure~\ref{fig:prior_comparison}),
and there are no data observed in the severity submodel,
i.e.~\(Y_{2} = \varnothing\). In stage one we target the
ICU submodel posterior \(\pd_{1}(\phi, \psi_{1} \mid Y_{1})\), enabling
the use of the original \texttt{JAGS} \citep{plummer:18} implementation.
These samples for \(\phi\) are displayed in the right panel of
Figure~\ref{fig:prior_comparison}, and we see that
whilst there is substantial overlap with \(\pd_{2}(\phi)\) (left panel),
\(\pd_{1}(\phi \mid Y_{1})\) is more disperse, particularly for
\(\phi_{1}\). Our region of interest is thus the HDR of
\(\pd_{1}(\phi \mid Y_{1})\), as the two-stage
sampler involves evaluating the samples from \(\pd_{1}(\phi \mid Y_{1})\)
under \(\pd_{2}(\phi)\).

\subsection{Self-density ratio estimation}
\label{sec:estimating-the-severity-self-density-ratio-for-stage-two}

The stage two acceptance probability for a move from
\(\phi \rightarrow \phi^{*}\) where
\(\phi^{*} \sim \pd_{1}(\phi, \psi_{1} \mid Y_{1})\) is
\begin{equation}
  \alpha(\phi^{*}, \phi) = \frac {
    \pd_{\text{pool}}(\phi^{*})
    \pd_{2}(\phi^{*}, \psi_{2}, Y_{2})
    \pd_{1}(\phi)
    \pd_{2}(\phi)
  } {
    \pd_{\text{pool}}(\phi)
    \pd_{2}(\phi, \psi_{2}, Y_{2})
    \pd_{1}(\phi^{*})
    \pd_{2}(\phi^{*})
  }.
  \label{eqn:icu-stage-two-acceptance-probability}
\end{equation}
In both the severity and ICU submodels, the prior marginal distribution
\(\pd_{\modelindex}(\phi)\) is unknown. This necessitates estimating the
self-density ratio for both \(\pd_{1}(\phi)\) and \(\pd_{2}(\phi)\).
However, the uniformity of \(\pd_{1}(\phi)\) corresponds to a
self-density ratio that is effectively \(1\) everywhere. In contrast,
the severity submodel prior marginal \(\pd_{2}(\phi)\) is clearly not
uniform over our region of interest; appropriately estimating the melded
posterior thus requires an accurate estimate of
\(\pd_{2}(\phi) \mathop{/} \pd_{2}(\phi^{*})\).

To obtain the WSRE estimate of $\pd_{2}(\phi) \mathop{/} \pd_{2}(\phi^{*})$ we first note that, in the notation of Section~\ref{sec:choice-of-xi}, $\phi = (\phi_{1}, \phi_{2}) = (\phi^{[1]}, \phi^{[2]})$, thus $D = 2$.
We set $V = 10$, resulting in $W = 10^{2}$ weighting functions, and we draw $10^{3}$ samples from each weighted target for a total of $10^{5}$ MCMC samples.
The Gaussian weighting functions for the first component $\m(\phi^{[1]}; \xi_{v, 1})$ have the variance parameter fixed to $25^{2}$; for the mean parameter we set $\xi_{1, 1} = 30$, $\xi_{V, 1} = 275$, with 8 other equally spaced values between these extrema.
For $\m(\phi^{[2]}; \xi_{v, 2})$ we set the variance parameter to $250^2$, with $\xi_{1, 2} = 500$, $\xi_{V, 2} = 3000$, and interpolate 8 equally spaced values between the extrema.
The values for the variance parameters are based on the empirical variance of the samples in Figure \ref{fig:prior_comparison}.

\subsection{Results}
\label{stage-two-trace-plots}

\begin{figure}
  \centering
  \includegraphics[width=\linewidth]{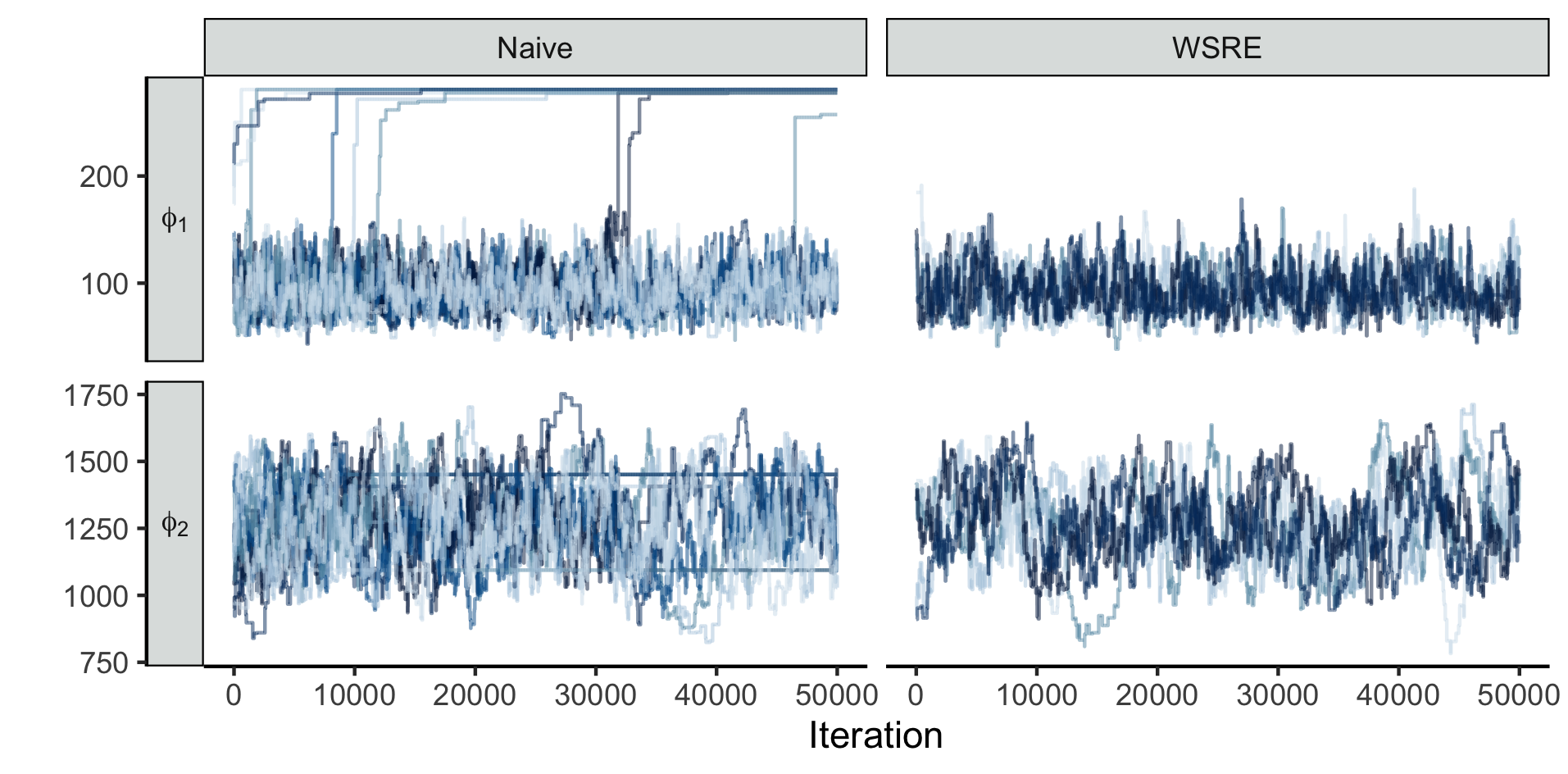}
\caption{Trace plots of 15 replicate stage two chains for $\phi_{1}$ and $\phi_{2}$, using the naive approach (left column) and the WSRE approach (right column).}
\label{fig:stage_two_phi_traces}
\end{figure}

We compare the melded posterior estimate obtained using WSRE against the naive approach.
For the latter we draw \(10^5\) samples from \(\pd_{2}(\phi)\) so that
both approaches have the same number of samples, although the naive
approach has a larger effective sample size.
Figure~\ref{fig:stage_two_phi_traces} displays trace
plots of 15 stage two MCMC chains, where \(\alpha(\phi^{*}, \phi)\) is
computed using the naive approach (left column), and using WSRE (right
column). The erroneous behaviour displayed in the left column is due to
underestimation of the tails of \(\widehat{\pd}_{2}(\phi)\) using a
standard KDE. This underestimation results in an overestimation of the
acceptance probability for proposals in the tails of
\(\widehat{\pd}_{2}(\phi)\), since the proposal term
\(\widehat{\pd}_{2}(\phi^{*})\) is in the denominator of
Equation~\eqref{eqn:icu-stage-two-acceptance-probability}.
Hence, moves to improbable values of \(\phi^{*}\) have acceptance
probabilities that are dominated by Monte Carlo error. Once at this
improbable value the error then has the opposite effect; the
underestimate yields chains unable to move back to probable values. This
produces the monotonic step-like behaviour seen in the top left panel of
Figure~\ref{fig:stage_two_phi_traces}. Although this
behaviour is not visible in all 15 chains, it will eventually occur if
the chains are run for more iterations, as a sufficiently improbable
value for \(\phi^{*}\) will be proposed. The results from this sampler
are thus unstable.

Whilst there is no baseline ``truth'' to compare to in this example, the
sampler that employs $\doublehat{\pdr}_{\text{WSRE}}(\phi, \phi^{*})$ as an estimate of \(\pd_{2}(\phi) \mathop{/} \pd_{2}(\phi^{*})\)
produces plausible results, in contrast to the naive approach. No
step-like behaviour is visible when employing the WSRE approach (right
column of ~\ref{fig:stage_two_phi_traces}). Whilst the between-chain mixing
is not optimal, this can be ameliorated by running the chains for longer, which
cannot be said for the naive method. This improved behaviour is obtained using
the same number of samples from the prior marginal distribution, or weighted
versions thereof. Users of this algorithm can be much more confident that the
results are not artefactual.

\section{Discussion}
\label{sec:discussion}

The complexity of many phenomena necessitates intricate,
large models. Markov melding allows the practitioner to channel
modelling efforts into smaller, simpler submodels, each of which may
have data associated with it, then coherently combine these smaller
models and disparate data. Multi-stage, sequential sampling methods,
such as the sampler used for Markov melding, are important tools for
estimating these models in a pragmatic, computationally feasible manner.

In particular, when an analytic form of the prior marginal distribution is not
available, we have demonstrated that the two-stage sampling process is
particularly sensitive to the corresponding KDE in regions of low probability. Tail
probability estimation is an important and recurrent challenge in
statistics \citep{hill:75,beranger_tail_2019}. We addressed this issue in
the Markov melding context by noting that we can limit our focus to the
self-density ratio estimate, and sample weighted distributions to
improve performance in low probability areas, for lower computational
cost than simple Monte Carlo. Our examples show that for equivalent
sample sizes, we improve the estimation of the melded posterior compared
to the naive approach.

The issue addressed in this paper arises to due differences in the
intermediary distributions of the two-stage sampling process,
particularly where the proposal distribution is wider than the target
distribution. The presence or absence of this issue is dependent upon
the order in which the components of the melded model are considered in the sampling process,
which is often constrained by the link function used to define \(\phi\)
in each model. In both our examples the link function is non-invertible.
\citet{goudie:etal:18} show extensions of the link function that render
it invertible are valid; that is, the model is theoretically invariant
to the choice of extension. However, the practical performance of the
two-stage sampler is heavily dependent on the appropriateness of such
extensions, and designing such extensions is extremely challenging.
Hence, the ordering of the submodels in the two-stage sampler is often
predetermined; we are practically constrained by the non-invertible link
function. In our examples this corresponds to sampling the less
informative model for \(\phi\) first. If we are free to choose the
ordering of the two-stage sampler, we may still prefer to sample the
wider model first, as the melded posterior is more likely to be captured
in a reweighted sample from a wider distribution than such a sample from
a narrow distribution. However, if the melded posterior distribution is
substantially narrower than the stage-one target distribution then we
are susceptible to the sample degeneracy and impoverishment problem
\citep{li:etal:14}. Addressing this issue in the melding context, whilst
retaining the computational advantages of the two-stage sampler, is an
avenue for future work.

The examples we consider contain 1 or 2 dimensional \(\phi\).
For higher dimensional \(\phi\) we anticipate encountering issues
associated with the curse of dimensionality. Specifically, the decrease
in accuracy of any KDE and increase in the required number of weighting
functions will scale exponentially with dimension.
Applying the argument in Section 3.4 to locate these additional weighting functions will be challenging.
As such we recommend
WSRE, like other KDE methods, for settings where
\(\text{dim}(\phi) \leq 5\) \citep{wand:jones:95}. This requirement may
be relaxed when there is structure in \(\phi\) that allows it to be
split into lower-dimensional components, such as when \(\phi\) contains
a collection of subject-specific parameters that are independent a
priori. More generally, in high dimensions almost everywhere is a
`region of low probability' and the performance of KDEs is known to be
poor, making choosing both an appropriate number of weighting functions
and their parameters difficult. Machine learning methods have proven to
be effective for estimating densities of moderate to high dimension (see
\citet{wang:scott:19} for a review), however the performance of these
methods in low probability regions has not, to our knowledge, been
thoroughly investigated.

There are potential alternatives to our weighted-sample self-density
ratio estimation technique. Umbrella sampling \citep{torrie_nonphysical_1977-1, matthews:etal:18}
aims to accurately estimate the tails of a density \(\pd(\phi)\) by
constructing an estimate \(\widehat{\pd}(\phi)\) from \(\Nw\) sets of
weighted samples
\(\{\phi_{\sampleindex, \wfindex} \}_{\sampleindex = 1}^{\Nx} \sim \s_{\wfindex}(\phi; \xi_{\wfindex})\),
However, umbrella sampling requires estimates of the normalising
constants \(Z_{2, \wfindex} = \int \s_{\wfindex}(\phi; \xi_{\wfindex}) \text{d}\phi\) to
combine the density estimates computed from each weighted sample. Our
approach is able to avoid computing normalising constants by focusing on
the self-density ratio.
Umbrella sampling also requires choosing the location of the weighting functions, i.e. choosing $\xi_{\wfindex}$ appropriately.
A heuristic strategy, similar to that of Section \ref{sec:choice-of-xi}, is seen as necessary by \citet{torrie_nonphysical_1977-1}.
Adaptive procedures that automatically choose values of $\xi_{\wfindex}$ based on other criteria exist, but these assume that $\s_{\wfindex}(\phi)$ is a Gaussian distribution \citep{mitsuta_automated_2018} or operate on a predefined grid of possible values \citep{wojtas-niziurski_self-learning_2013}.
We cannot use the generic tempering methodology advocated by \citet{matthews:etal:18}, as sampling from $\pd(\phi, \gamma)^{1 \mathop{/} \tau}$, for $\tau > 1$, does not generally produce marginal samples from $\pd(\phi)^{1 \mathop{/} \tau}$.

Another possibility would be to sample \(\pd_{\text{meld}}\) using a
pseudo-marginal approach \citep{andrieu:roberts:09}. A necessary
condition of the pseudo-marginal approach is that we possess an unbiased
estimate of the target distribution. Kernel density estimation produces
biased estimates of \(\pd(\phi)\) for finite \(\Nx\). A KDE can be
debiased \citep{calonico:cattaneo:farrell:18, cheng:chen:19}, but doing
so requires substantial computational effort. Moreover, we also require
an unbiased estimate of \(1 \mathop{/} \pd(\phi)\). Debiasing estimates
of \(1 \mathop{/} \pd(\phi)\) is possible with pseudo-marginal methods
like Russian roulette \citep{lyne:etal:15}, but \citet{park:haran:18}
observe prohibitive computational costs when doing so. The presence of
both \(\ppoolphi\) and \(1 \mathop{/} \pd(\phi)\) in the melded
posterior further complicates the production of an unbiased estimate,
particularly when \(\ppoolphi\) is formed via logarithmic pooling.

\section*{Acknowledgements}

We acknowledge Anne Presanis for many helpful discussions about this issue, and thank the referees for useful comments on an earlier version of this paper.
This work was supported by The Alan Turing Institute under the UK Engineering and Physical Sciences Research Council (EPSRC) [EP/N510129/1] and the UK Medical Research Council [programme code MC\_UU\_00002/2].

\bibliographystyle{spcustom}
\bibliography{year1bib}

\end{document}